\documentclass[aps,prl,reprint,superscriptaddress,longbibliography]{revtex4-2}

\usepackage{graphicx,hyperref}
\usepackage{amsmath}
\usepackage{amsfonts}
\usepackage{epsfig} 
\usepackage{color}
\usepackage{bbm}
\usepackage{amsthm}

\newcommand{\beq}{\begin{equation}}
\newcommand{\eeq}{\end{equation}}
\newcommand{\bea}{\begin{eqnarray}}
\newcommand{\eea}{\end{eqnarray}}

\mathchardef\nss="711B

\def\nss{\mathcal{S}}

\def\be{\begin{eqnarray}}
\def\ee{\end{eqnarray}}

\newlength{\myL}

\begin{document}

\title{Non-Gaussian diffusive fluctuations in Dirac fluids}

\author{Sarang Gopalakrishnan}
\affiliation{Department of Electrical and Computer Engineering, Princeton University, Princeton NJ 08544}

\author{Ewan McCulloch}
\affiliation{Department of Electrical and Computer Engineering, Princeton University, Princeton NJ 08544}
\affiliation{Department of Physics, University of Massachusetts, Amherst MA 01003}

\author{Romain Vasseur}
\affiliation{Department of Physics, University of Massachusetts, Amherst MA 01003}

\begin{abstract}
Dirac fluids---interacting systems obeying particle-hole symmetry and Lorentz invariance---are among the simplest hydrodynamic systems; they have also been studied as effective descriptions of transport in strongly interacting Dirac semimetals. Direct experimental signatures of the Dirac fluid are elusive, as its charge transport is diffusive as in conventional metals. In this paper we point out a striking consequence of \emph{fluctuating} relativistic hydrodynamics: the full counting statistics (FCS) of charge transport is highly non-gaussian. We predict the exact asymptotic form of the FCS, which generalizes a result previously derived for certain interacting integrable systems. A consequence is that, starting from quasi-one dimensional nonequilibrium initial conditions, charge noise in the hydrodynamic regime is parametrically enhanced relative to that in conventional diffusive metals.
\end{abstract}

\maketitle

Hydrodynamics is an effective theory for the coarse-grained non-equilibrium dynamics of interacting systems~\cite{landau1987fluid,Chaikin_Lubensky_1995}. Hydrodynamics assumes that all but a few degrees of freedom relax fast. Symmetries and conservation laws determine the remaining slow modes, their equations of motion, and consequently, e.g., the dynamical exponent that governs how perturbations spread. Hydrodynamics has usually been interpreted as a theory of \emph{expectation values}; however, it can also be extended to a framework (``fluctuating hydrodynamics'') that describes the spatio-temporal fluctuations of conserved densities~\cite{forster2018hydrodynamic, spohn2012large, liu2018lectures}. Such density fluctuations are accessible in present-day experiments with cold atoms and superconducting qubit arrays~\cite{gross2017quantum, arute2019quantum, scholl2021microwave, PhysRevA.95.053621, hofferberth2008probing, kitagawa2011dynamics, chiu2019string, rosenberg2023dynamics, 2016Sci...353.1257B, 2020PhRvL.125a0403K, 2022PhRvL.129l3201Y, 2023Sci...381...82H,wei2022quantum, wienand2023emergence,2016Sci...353..794K,2023arXiv231213268I}, and are related to nonlinear response coefficients that can be probed using pump-probe spectroscopy~\cite{mahmood2021observation, fava2021hydrodynamic, delacretaz2023nonlinear}. 
A central question in fluctuating hydrodynamics is whether density fluctuations obey universal scaling laws, and if so, what determines their universality classes. 

In the present work, we explore this question for ``Dirac fluids,'' i.e., Lorentz-invariant systems with particle-hole symmetry. Lorentz-invariant fluids \cite{KovtunNotes} have recently been considered in many contexts including heavy-ion collisions \cite{heavyionlecturenotes,heavyionreview} and, most prominently, the dynamics of electrons in semimetals such as graphene \cite{Muller2008,Lucas2016a,Lucas2016b,Crossno2016,Narozhny2017,Lucas2018,Bal2021}. In solid-state systems, the crystal lattice breaks Lorentz invariance through umklapp scattering, but this scattering rate is negligible in the cleanest available samples \cite{Bandurin2016,Moll2016,KrishnaKumar2017,Lin2019}. Nevertheless, finding unambiguous signatures of hydrodynamic behavior has been a challenge: 
while heat transport in Dirac fluids is ballistic as the energy current is proportional to momentum density, the charge current can decay because of particle-hole symmetry. The hydrodynamic theory thus predicts that charge transport is simply diffusive, so experimental efforts have had to study heat transport (which strongly violates the Wiedemann-Franz law) and thermoelectric effects~\cite{PhysRevB.79.085415,Fong2012,Fong2013,Crossno2016,Ghahari2016}, which are challenging to detect. 

\begin{figure}[!b]
    \centering
    \includegraphics[width=0.42\textwidth]{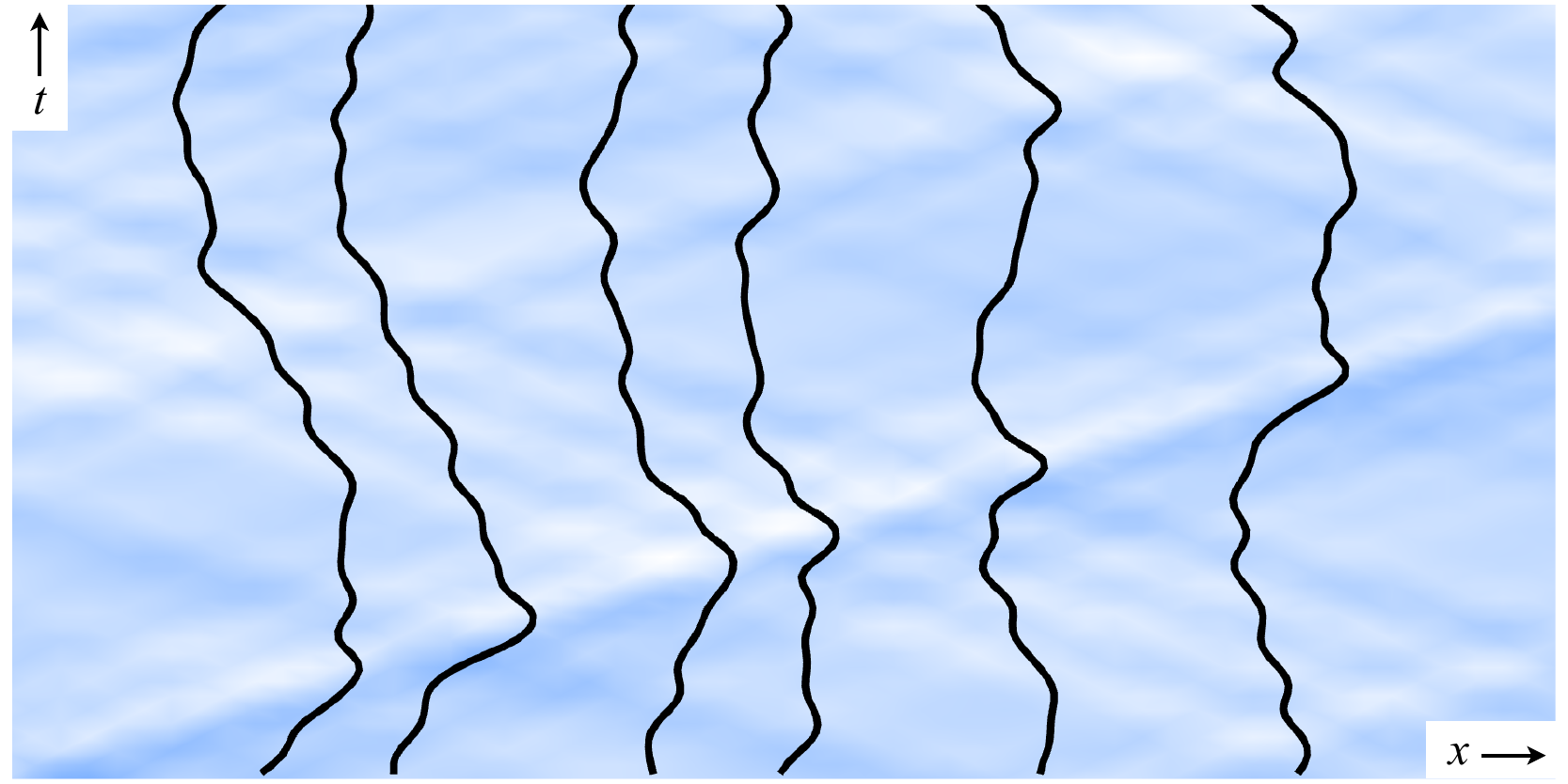}
    \caption{{\bf Diffusion from convection.} Schematic depiction of the convective mechanism for charge diffusion. Thermal energy fluctuations in the initial state (shades of blue) propagate as sound waves that impart fluctuating velocities to representative charge elements (black). Each charge element undergoes a random walk, but the random walks of different charge elements are correlated: all of them encounter the same sound waves.}
    \label{fig1}
\end{figure}

We find that charge noise in the Dirac fluid is anomalous, both in and out of equilibrium. For concreteness, we characterize the noise through the full counting statistics (FCS) of charge transport in quasi-one dimensional geometries~\cite{Levitov1993, Belzig2001, Levitov_2004, RevModPhys.81.1665}, but the mechanism we consider extends to general fluctuation phenomena in finite-temperature states (or nonequilibrium states at finite entropy density). This mechanism is closely related to the one recently identified in integrable systems \cite{anomalousFCS_spinchain_review,Gopalakrishnan2022a,Krajnik2022a,Krajnik2022b,Krajnik2022c,Krajnik2023a} (see also~\cite{Myers2018, doyon2020fluctuations, PhysRevLett.131.027101, doyon2023ballistic}). In both the integrable and relativistic cases, diffusion is due to the interaction of slow collective charged modes with ballistically propagating neutral modes~\cite{PhysRevLett.119.080602,Gopalakrishnan2018,Medenjak2020,Doyon2022}. Because the neutral modes are long-lived, the noise they generate has strong spatio-temporal correlations, so the charge dynamics is that of a few giant diffusing objects rather than many independently moving diffusive particles, and the noise is strongly enhanced (Fig.~\ref{fig1}). Out of equilibrium, charge fluctuations across a cut are parametrically enhanced in the Dirac fluid relative to a conventional diffusive system. In equilibrium, their variance is set by the fluctuation-dissipation theorem; however, the higher moments of noise remain anomalous {-- in the sense that they violate expectations from large deviations and central limit theorem arguments}. We show that these features are a general consequence of a simple three-mode fluctuating hydrodynamics that emerges in both the relativistic and integrable contexts. 

We will primarily be interested in the FCS of charge in a macroscopic subsystem. The FCS can be defined and measured in many equivalent ways; for conceptual simplicity, we define it through the following protocol. On a given run of the experiment the subsystem of interest, $S$, is initialized in a state of definite charge $Q_0$. It is then suddenly connected to the rest of the system and the total charge in $S$ is measured after some time $t$, giving a random outcome $Q$ from a distribution $P_t(Q)$. We are interested in $P_t(\Delta Q = Q - Q_0)$. 
The theory of FCS in generic diffusive systems is ``macroscopic fluctuation theory''\cite{Bertini_2015}, which posits that this entire distribution is determined by the dependence of the diffusion constant on the conserved densities in the system. In standard metals, this dependence is smooth, giving rise to conventional asymptotically gaussian FCS. In relativistic hydrodynamics, however, this dependence is singular~\cite{Lucas2018} (as the diffusion constant is finite at charge neutrality, but diverges away from it), so the FCS \emph{cannot} be conventional. 

\emph{Three-mode hydrodynamics}.---The hydrodynamics we will consider has three conserved densities: charge $n$, energy $e$, and energy current $\phi_i$ ($i$ runs over spatial indices, and we implicitly sum over repeated indices). 
$\phi_i$ might be conserved as a consequence of Lorentz invariance or for any other reason. The first hydrodynamic equation is essentially a definition of $\phi_i$ being the energy current:
\beq\label{cont_energy}
\partial_t e + \partial_i \phi_i = 0.
\eeq
Next we write a continuity equation for $\phi_i$. In what follows we will consider $n$, $e$ and $\phi_i$ to be small deviations from an equilibrium background, and write down linearized hydrodynamic equations, keeping nonlinearities only when important (see below for justification). Since $\phi_i$ is already a vector its current is a second-rank tensor $q_{ij}$, which is even under time-reversal and inversion and neutral under the conserved charge. The tensor $e \delta_{ij}$ satisfies the necessary symmetries to appear in $q_{ij}$, up to an order one constant that we set to unity for simplicity. Ignoring any other potential conserved quantities 
%{\color{red} and nonlinear terms such as $\phi_i \phi_j$ that will not be important to us}, 
we arrive at the equation
\beq\label{cont_phi}
\partial_t \phi_i + \partial_i e + \cdots = 0,
\eeq
where $\cdots$ consists of terms beyond Euler scale that will not be relevant. In the relativistic context, the energy current $\phi_i$ is also momentum density (by symmetry of the stress energy tensor $T_{\mu \nu}$), and the current of energy current is pressure, which is indeed directly proportional to energy density from the equation of state -- see {\it e.g.}~\cite{Lucas2018, BhaseenRelativistic, PhysRevD.94.025004, Vasseur2015}. Equations~\eqref{cont_energy} and~\eqref{cont_phi} imply that both $e$ and $\phi_i$ propagate ballistically,   
\beq
\partial^2_t e - \partial_i\partial_i e  = 0,
\eeq
Divergenceless patterns of the velocity field $\partial_i \phi_i=0$ are frozen in place at the Euler scale, but this will not play any role in the quasi-one dimensional geometries considered in this letter. 
%Upon including higher-order gradient corrections (beyond Euler), the ballistic ``sound waves'' carried by $\phi_i$ will be broadened, either diffusively in dimensions $d>2$, or with anomalous (KPZ) scaling in $d=1$ due to nonlinear contributions~\cite{PhysRevLett.56.889, PhysRevLett.108.180601, SpohnNLFH,PhysRevLett.54.2026, PhysRevA.92.043612,Fibonacci}. This broadening will be unimportant for our purposes.

It remains to write down a continuity equation for $n$. At charge neutrality the equations have a particle-hole symmetry $n \to -n$, which prevents the charge current from overlapping with energy current fluctuations. Within \emph{linearized} hydrodynamics, charge transport is purely diffusive at this point, with a diffusion constant $D$ and a corresponding (conserving) noise term $\partial_i \xi_i$.
Away from charge neutrality---i.e., at net charge $n_0$---charge and energy fluctuations mix; the charge current has a term $\sim n_0 \phi_i$, and charge transport is ballistic. 
Even at charge neutrality the charge has local fluctuations $n$, which couple to energy current, giving a nonlinear coupling $\sim n \phi_i$, again up to an irrelevant proportionality constant that we set to unity. 
%{\color{red} It is well-known that such nonlinear terms can drastically change hydrodynamic behavior, giving rise to long-time tails and frequency-dependent hydrodynamic coefficients~\footnote{Those renormalization effects will not be important in the 1d geometry that we consider: the charge mode remains diffusive as it cannot interact with itself (from particle-hole symmetry) or with shear modes that are absent in 1d. Away from charge neutrality, the putative  diffusive mode becomes superdiffusive with $z=5/3$~\cite{PhysRevLett.108.180601,PhysRevLett.89.200601,SpohnNLFH,Fibonacci}. }. For our purposes of computing FCS, the effects of the nonlinear term will be more subtle but still qualitatively important.} 
Including both the diffusive and nonlinear terms we arrive at the equation
%Under particle-hole symmetry, $n \to -n, e \to e, \phi_i \to \phi_i$; this symmetry holds both in charge-neutral graphene and in the easy-axis XXZ model**, and prevents charge fluctuations from overlapping with energy fluctuations above such symmetric states. Within \emph{linearized} hydrodynamics, charge transport is purely diffusive at this point, with a diffusion constant $D$ and a corresponding noise term $\xi$.
%Away from charge neutrality---i.e., at net charge $n_0$---charge and energy fluctuations mix; the charge current has a term $\sim n_0 \phi_i$. 
%Under this assumption, the most general hydrodynamic equation one can write for $n$ is
\beq\label{cont_charge}
\partial_t n + \partial_i (\phi_i n) = D \partial_i \partial_i n + \partial_i\xi_i.
\eeq
%
% It remains to write down a continuity equation for $\phi_i$. Since $\phi_i$ is already a vector its current is a second-rank tensor $q_{ij}$, which is even under time-reversal and inversion. Crucially, the tensor $e \delta_{ij}$ satisfies the necessary symmetries to appear in $q_{ij}$. Ignoring any other potential conserved quantities, we arrive at the equation
% \beq\label{cont_phi}
% \partial_t \phi_i + \partial_i e  = 0.
% \eeq
Eqs.~\eqref{cont_energy}, \eqref{cont_phi}, and \eqref{cont_charge} define a closed three-mode non-linear fluctuating hydrodynamic system, which coincides in the relativistic case with the equations derived in Ref.~\cite{Lucas2018} in the linear case. 
As discussed above, the equations for $e$ and $\phi_i$ can be combined into a wave equation. Energy and charge fluctuations travel at different speeds and decorrelate. At each time an energy (charge) fluctuation sees a totally different charge (energy) environment. Since we are interested in charge transport, this decoupling justifies leaving Eq.~\eqref{cont_energy}, \eqref{cont_phi} at Euler scale. 

\emph{Nonlinear terms and fluctuations}.---Above, we wrote a very simple set of hydrodynamic equations, ignoring various nonlinearities that are known to qualitatively affect low-dimensional hydrodynamics, giving rise to long-time tails and frequency-dependent hydrodynamic coefficients. We now briefly justify this neglect. First, the current of energy current~\eqref{cont_phi} contains nonlinear terms of the form $\phi_i \phi_j$ by symmetry. Upon including higher order gradient corrections, the ballistic ``sound waves'' carried by $\phi_i$ will be broadened, either diffusively in dimensions $d>2$, or with anomalous (KPZ) scaling in $d=1$ due to these nonlinear contributions~\cite{PhysRevLett.56.889, PhysRevLett.108.180601, SpohnNLFH,PhysRevLett.54.2026, PhysRevA.92.043612,Fibonacci}. This broadening is subleading to the Euler-scale propagation of sound waves and can therefore be neglected for our purposes. Second, the charge current~\eqref{cont_charge} contains a nonlinear coupling to the energy current---whose effects on the FCS we \emph{will} keep. One might worry however that this nonlinear coupling will make the charge spread superdiffusively: for example, in one-dimensional Galilean fluids the ``heat mode'' is diffusive in linearized hydrodynamics but spreads as $t^{3/5}$ once nonlinearities are included~\cite{PhysRevLett.108.180601,PhysRevLett.89.200601,SpohnNLFH,Fibonacci}. 
In the present case, particle-hole symmetry prevents the charge mode from having either a KPZ nonlinearity or a $t^{3/5}$ nonlinearity. Finally, shear modes that the charged mode could couple to do not exist in one dimension. 
Therefore, in one-dimensional fluids with particle-hole symmetry, diffusive hydrodynamics is stable at the level of equilibrium correlators and the leading nontrivial effect is on the FCS, as we now discuss.

\begin{figure*}[t]
  \includegraphics[width=0.34\textwidth]{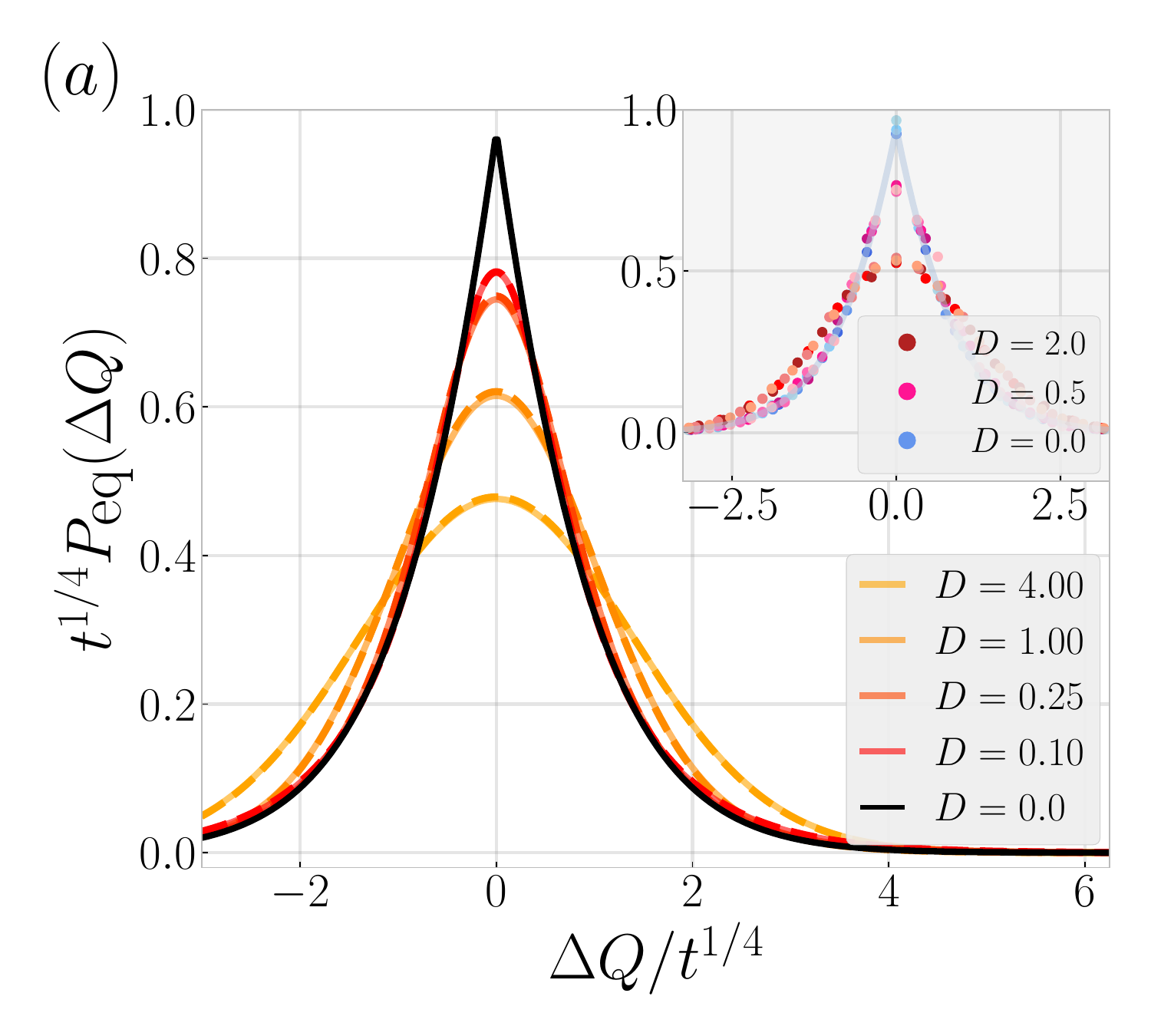}
  \hspace{-3mm}\includegraphics[width=0.34\textwidth]{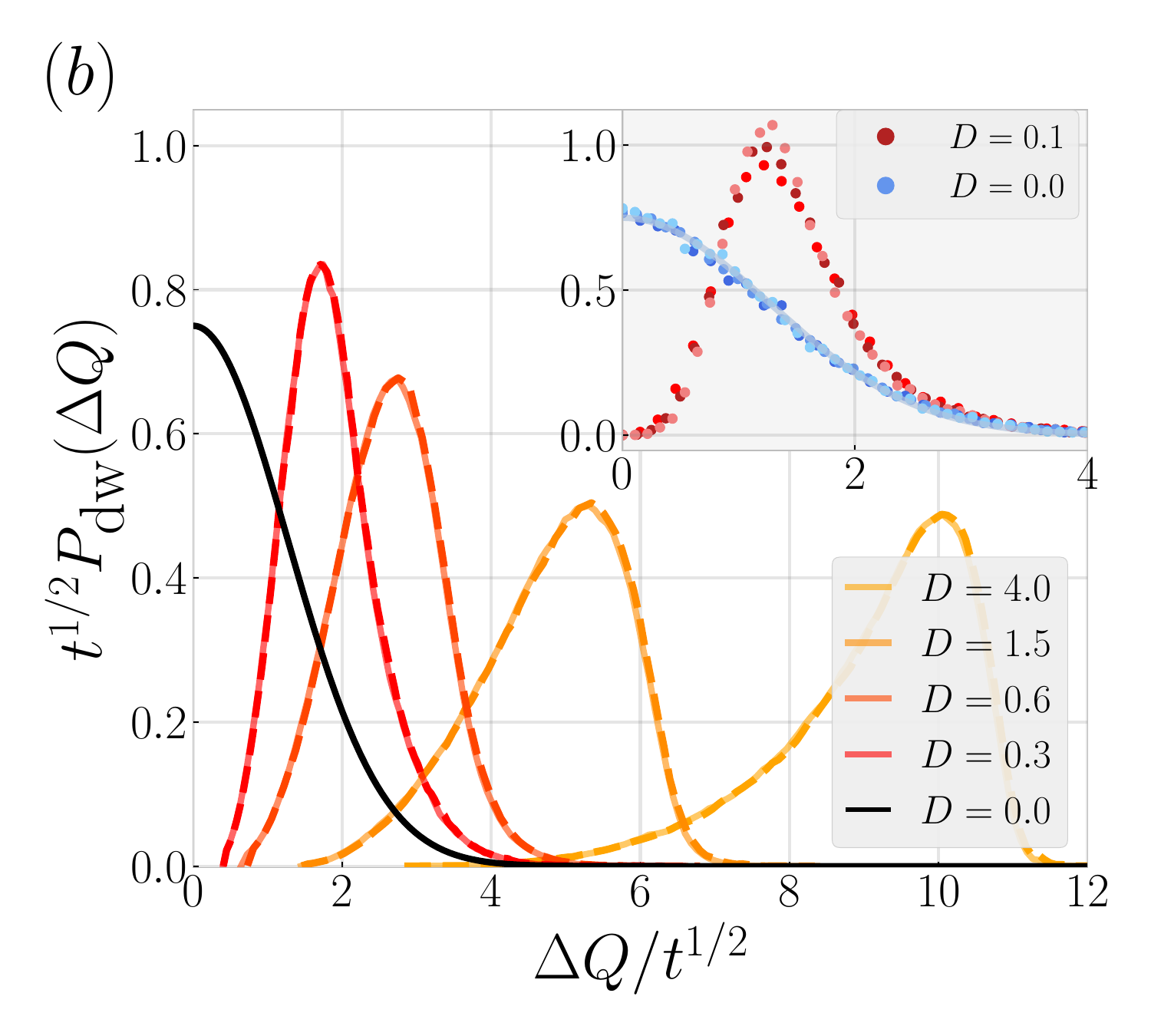}
  \hspace{-3mm}\includegraphics[width=0.34\textwidth]{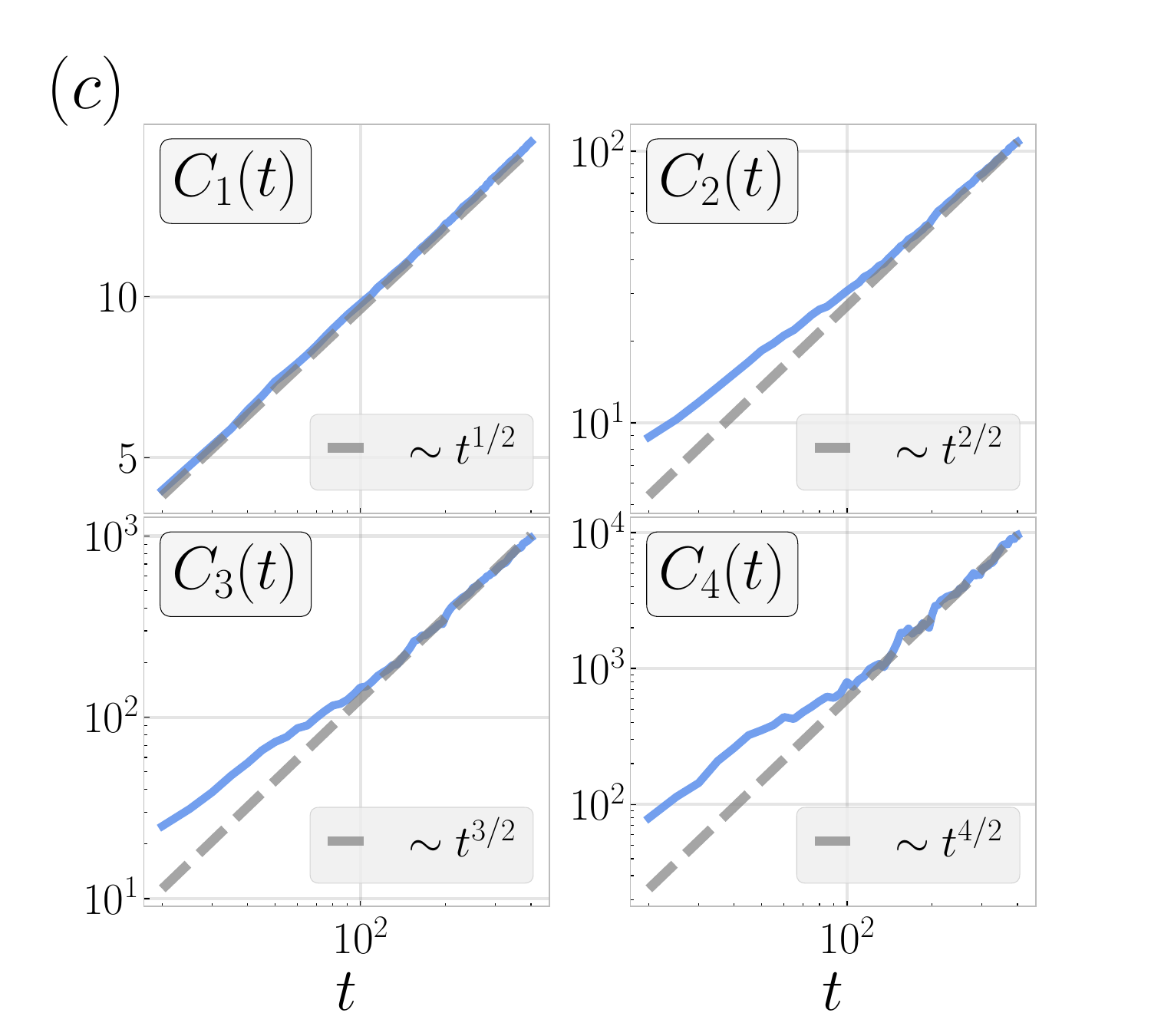}
  \caption{{\bf Anomalous full counting statistics.} Charge transfer distributions $P(\Delta Q(t))$ in one dimension: (a) in equilibrium and (b) for a domain wall initial state. Charge transfer is calculated approximately, by evaluating the charge transfer integral (Eq.~\ref{dw transfer int} in the domain wall case) in the absence of the noise $\xi$ (panels (a) and (b)) for various diffusivities $D$, and directly, by numerically solving equation~(\ref{cont_charge}) (insets (a) and (b)). 
  %In the $D=0$ limit, and within the self consistent approximation, Eq.~(\ref{cont_charge}) can be solved by the method of characteristics and yields distributions with non-analyticities at $\Delta Q = 0$. 
  Ordinary diffusion smooths the non-analyticities at $\Delta Q = 0$, but the distributions remain highly non-Gaussian. A scaling collapse of $P_t(\Delta Q)$ is observed under the rescaling $\Delta Q/t^{1/4}$ in equilibrium and $\Delta Q/t^{1/2}$ for a domain wall; the scaled distributions are shown for two times $t=400$ (dashed) and $t=800$ (solid) in the main panels and for multiple times $t=400,600,800$ (multiple shades) in the insets. The cumulants of charge transfer in the domain wall case asymptotically scale as $C_n(t)\sim t^{n/2}$ (shown in panel (c)), indicating anomalous transport statistics.}\label{fig2}
\end{figure*}

\emph{Charge diffusion by convection}.---Eq.~\eqref{cont_charge} describes two qualitatively different diffusion mechanisms. The first is standard diffusion, described by the right hand side. The second is charge diffusion due to energy current fluctuations: because of hydrodynamic decoupling, $\phi_i$ is effectively a random (from thermal fluctuations in the initial state), temporally uncorrelated velocity for charge fluctuations, and a particle evolving with a random velocity undergoes diffusion.
This nonlinear diffusion mechanism has recently been explored extensively in the integrability literature~\cite{PhysRevLett.119.080602,2018PhRvL.121p0603D,Gopalakrishnan2018,DeNardis2019,Medenjak2020,Doyon2022, Gopalakrishnan2022a,anomalousFCS_spinchain_review,Krajnik2022a,Krajnik2022b,Krajnik2022c,Krajnik2023a}, and is the source of the anomalous fluctuations discussed here. 

%Note that in these equations the dynamics of charge does not feed back into that of energy: rather, slowly moving charge fluctuations diffuse as a consequence of being buffeted by charge-neutral sound waves. 

\emph{Toy model}.--- For simplicity we will first ignore the ``standard'' diffusion term, and reinstate it later; and also consider the case of one spatial dimension $d=1$. Charge transport is then described by the equation
\beq\label{cont_charge_simplified}
\partial_t n + \partial_x (\phi n) = 0,
\eeq
where $\phi$ effectively decouples and becomes a source of noise for charge $n$. It remains however a specific type of noise, as it propagates ballistically $\phi (x,t)  = \phi_R (x- v t) +  \phi_L (x+ v t)$, with $v$ the sound velocity, up to (KPZ, in 1d~\cite{PhysRevLett.56.889, PhysRevLett.108.180601, SpohnNLFH,PhysRevLett.54.2026, PhysRevA.92.043612,Fibonacci}) broadening corrections. As we now discuss, charges moving with noisy but ballistically-correlated velocities undergo some diffusive motion but with anomalous fluctuations and full counting statistics. 

To see this, we first note that eq.~\eqref{cont_charge_simplified} can be solved using the characteristics equation
\begin{align}
\partial_\tau X & = \phi(X(x,t,\tau), \tau), \notag\\ 
X(x,t,t) & = x. 
\end{align}
Now proceeding self-consistently, the noisy velocity field $\phi$ will lead to diffusive characteristics, so we expect $X(x,t,\tau) = x + {\cal O}(\sqrt{\tau-t})$ . Therefore, we can approximate $X(x,t,\tau) \pm v \tau \approx x \pm v \tau $ up to terms of order $\sqrt{\tau}$. We thus have $X(x,t,\tau) \approx x + \int_t^\tau d\tau \phi(x,\tau)$ at long times $\tau$ -- this allows us to evaluate $\phi(x,\tau)$ at position $x$ instead of $X(x,t,\tau)$. At long times, we find
\begin{equation} \label{eqHydrosolution}
n(x,t) \approx n_0 \left(x - \int_0^t d\tau \phi(x,\tau) \right).
\end{equation}
Let us now use this result to compute the FCS of charge of the right-half $x \geq 0$ of the system, $Q(t) = \int_0^\infty dx n(x,t)$. Using the continuity equation~\eqref{cont_charge_simplified}, we have $\dot Q = \phi(x=0,t) n(0,t) $. Integrating that equation and using~\eqref{eqHydrosolution}, we find:
\begin{equation} 
\Delta Q(t) = \int_0^{X(t)} d X_t n_0 \left( - X_t \right),
\end{equation}
where $X(t) = \int_0^t d\tau \phi(0,\tau)$ undergoes a random walk with a diffusion constant set by the variance of $\phi(0,\tau)$. 

\emph{Anomalous fluctuations}.--- The statistics of the charge transfer~\eqref{eqHydrosolution} is highly anomalous and different from that of ordinary chaotic diffusive systems. This is especially clear if we take the initial charge distribution to be a sharp domain wall $n_0(x) = - n_0$ if $x>0$, and $n_0(x) = + n_0$ for $x<0$. Then, $\Delta Q(t)$ is directly proportional to $|X(t)|$. The distribution $P_t(\Delta Q)$ is therefore highly skewed, and is that of a random walk on the half-line. In particular, while the mean charge transfer scales diffusively as expected $\langle \Delta Q (t) \rangle \sim \sqrt{t}$, its variance also scales as $\langle\Delta Q (t) ^2 \rangle_c \sim t$, in sharp contrast with the typical behavior expected of generic diffusive systems where all cumulants scale as $\sqrt{t}$ (consistent with expectations from central limiting behavior and large deviations). More generally, the full distribution $P_t(\Delta Q)$ can be obtained from rewriting~\eqref{eqHydrosolution} as $\Delta Q(t) = |X(t)| m_{[0,X(t)]} $ where $m_{[0,X(t)]}$ is the magnetization density in the interval between the origin and $X(t)$. Here, $X(t)$ is normally distributed with variance $2D_{X} t$ and zero mean, and in thermal equilibrium at the charge neutral point, $n_{[0,x]}$ is a normally distributed variable with zero mean and variance $T \chi(T)/|x|$ where $\chi = \left. d n/d \mu \right|_{\mu=0}$ is the charge susceptibility -- in what follows we will set $T \chi=1$ for simplicity. This gives the equilibrium FCS as a function of time
\begin{equation}
P_t(\Delta Q) = \int_0^\infty dX \frac{{\rm exp} \left( - \frac{X^2}{4 D_{X} t} - \frac{\Delta Q^2}{2 X} \right)}{\sqrt{2 \pi^2  } \sqrt{D_{X} t X}}.
\end{equation}
This formula was recently obtained in the context of spin transport in integrable XXZ chains~\cite{Gopalakrishnan2022a} (for which energy current is also conserved as a consequence of integrability), integrable classical spin-chains \cite{Krajnik2023a}, and of some integrable cellular automata~\cite{Krajnik2022a,Krajnik2022c}. This distribution characterizes charge transfer at the scale $\Delta Q \sim t^{1/4}$, and is highly non-Gaussian and non-analytic at $\Delta Q=0$. When the system is biased slightly out of equilibrium, with a domain wall of contrast $\epsilon$, the growth of variance crosses over from equilibrium to the nonequilibrium behavior on a timescale $t_\star(\epsilon) \sim 1/\epsilon^4$. 

\emph{Restoring ordinary diffusion}.---
We now consider the effects of ordinary diffusion on the charge transfer by restoring the Fick’s law term in Eq.~\ref{cont_charge}. First, we note that we can neglect the associated noise term $\xi$ in the case of the domain-wall distribution. Over a time $t$, the convective term moves the domain wall a distance $t^{1/2}$; meanwhile, the deterministic part of the Fick's-law diffusion smears the domain wall out over a distance $t^{1/2}$, and the associated noise contributes $t^{1/4}$ to the charge transfer variance. This last contribution is subleading for the domain-wall initial condition and we will neglect it.

%required by the fluctuation-dissipation theorem, as it does not turn out to be important for non-equilibrium initial distribution such as the domain wall distribution we consider now. In the frame of a domain wall undergoing a random walk $X(t)$, charge may transfer across some position $r$ through ordinary diffusion. This leads to an additional contribution to the mean charge transfer of $\mathcal{O}(\sqrt{t})$, whose fluctuations, at $\mathcal{O}(t^{1/4})$, are unimportant corrections to the charge transfer fluctuations from the random walk of the domain wall.

Proceeding without the noise term, Eq.~\ref{cont_charge} can be solved by convolving a solution $\tilde{n}$ of the toy model with the diffusion kernel $G$ for ordinary diffusion, $n(x,t) = (\tilde{n} * G)(x,t)$. Charge transfer now has two contributions, one from the shared kicks from $\phi$, and another from Fick's law: $\Delta Q(t) = \int_0^t d\tau \int dx\ n_0(x-X(\tau)) (\phi(\tau)+D\partial_{x}) G(x,\tau)$. For a domain-wall initial, this is once again given in terms of a random walk $X_t$,
\begin{align}\label{dw transfer int}
    \Delta Q(t) = n_0& \int_0^t d\tau \ \text{erf}\left(\frac{X(\tau)}{\sqrt{2D\tau}}\right)\frac{d X}{d \tau} \nonumber\\
    +&\ 2 n_0 D \int_0^t d\tau \ G(X(\tau),\tau).
\end{align}
By sampling over random walks $X_t$ (with $D_X=1$), we numerically evaluate this integral for various diffusivities $D$ and find the probability distribution $P_t(\Delta Q)$ (shown in Fig.~\ref{fig2} (b)). The inclusion of ordinary diffusion smooths the non-analyticity at $\Delta Q=0$, but the distribution remains non-Gaussian and is highly skewed at both small and large diffusivity. The distributions obey a scaling collapse with the rescaling $\Delta Q/\sqrt{t}$, and all cumulants scale anomalously as $C_n\sim t^{n/2}$ (see Fig.~\ref{fig2} (c)).

In equilibrium, the fluctuations associated with ordinary diffusion are a relevant correction to the charge transfer statistics and neglecting them will yield quantitatively inaccurate statistics. Nevertheless, by restoring deterministic diffusion, the cusp at $\Delta Q = 0$ is removed while the distribution remains non-Gaussian and is qualitatively similar to the distribution obtained by solving Eq.~\ref{cont_charge} numerically (see Fig.~\ref{fig2} (a) and inset).

%By defining the re-scaled variable $Y(t) \equiv X(t)/\sqrt{t}$ (and also a rescaled noise $u(t) \equiv \phi(t)/\sqrt{t}$), so that $Y$ also has unit variance

%That is, restoring ordinary diffusion has the effect of smearing the initial density profile with the diffusion kernel $G$ but does not change the anomalous scaling of the charge transfer cumulants. In equilibrium, 

%ordinary diffusion modifies the charge transfer distribution by softening the cusp at $\Delta Q(t) = 0$.

\emph{General geometries; experiments}.---So far, we have focused on quasi-one-dimensional geometries, for which it makes sense to define FCS in a segment of the system. These geometries can be much wider than an atomic spacing (e.g., wide enough that it is appropriate to use the bulk dispersion for graphene) provided that the segment in which FCS is evaluated is even larger than the transverse width $w$. In such geometries, at Euler scale the only truly slow modes are those carrying momentum along the infinitely long direction. By the central limit theorem the strength of the fluctuations discussed here scales down as $\sim 1/\sqrt{w}$, so narrower channels are better. %Quasi-1D geometries can be realized both in synthetic matter---e.g., interacting particles in optical lattices with honeycomb dispersion---and in solid-state settings using gate-defined channels to confine a 2D electron gas.
Quasi-1D geometries can be realized both in synthetic matter---e.g., interacting particles in optical lattices, where honeycomb dispersia have already been synthesized \cite{2012Natur.483..302T,2015Sci...347..288D,2016Sci...352.1091F,2016Sci...352.1094L,2023PhRvE.107d4213B}---and in solid-state settings using gate-defined channels to confine a 2D electron gas \cite{PhysRevLett.56.1198,PhysRevLett.57.1769,Berggren2010}.
In cold-atom systems, imaging with single-site resolution is possible, so the measurement of FCS is feasible and has been demonstrated \cite{wei2022quantum,wienand2023emergence,2016Sci...353..794K,2023arXiv231213268I}. In the solid-state setting, direct FCS measurements are harder, although they have been accomplished using quantum point contacts~\cite{gossard}. Moreover, recent experimental advances (e.g., noise spectroscopy using nitrogen-vacancy centers in diamond \cite{NVCSpectroscopyReview}) in principle allow for more robust measurements of bulk FCS involving larger subsystems.

In higher dimensions, FCS across macroscopic bipartitions is asymptotically gaussian because it is a sum over many independent regions. However, the basic physics we have outlined---of ballistic energy fluctuations creating correlated charge noise---can be seen more generally using nonlinear spectroscopy. Nonlinear charge response involves correlation functions like $\langle n^2(x,t) n^2(0,0) \rangle$, which spread ballistically because $n^2$ has the same symmetries as $e$ and can therefore mix with it. Thus the nonlinear charge response in the Dirac fluid is radically different from the response that was recently predicted for standard diffusive systems~\cite{2023arXiv230403236D}. %Devising specific near-term experimental probes of this effect is an important task for future work.

\emph{Discussion}.--- The key result of this letter is the non-Gaussian nature of diffusive charge fluctuations in relativistic hydrodynamics. Our predictions rely on the following ingredients: (1) the existence of a conserved energy current, (2) particle-hole symmetry, (3) thermal fluctuations in the initial state, and (4) the absence of other conserved quantities that would overlap with the charge current, so charge transport is diffusive. Those conditions are satisfied in Dirac fluids (in particular, (1) follows from Lorentz invariance), but also in the gapped phase of the XXZ quantum spin chains~\cite{Gopalakrishnan2022a} and in some integrable cellular automata~\cite{Krajnik2022a,Krajnik2022c} -- in those integrable examples, the ordinary diffusion contribution is equal to zero. 
A natural question is whether our results would carry over to  energy transport in {\em Galilean invariant} systems, where charge current (momentum) is conserved (instead of energy current in Lorentz invariant systems). However, the analog of particle-hole symmetry in that case $e \to -e$ only holds at infinite temperature, ruling out continuum models: in general, energy in Galilean-invariant systems propagates ballistically through sound waves. 
%In addition, the diffusive ``heat mode'' in such systems broadens anomalously (not diffusively) in one dimension~\cite{PhysRevLett.108.180601,PhysRevLett.89.200601,SpohnNLFH,Fibonacci}. 

Our results show that, although charge transport in the Dirac fluid is diffusive at the level of \emph{averages}, the associated noise is qualitatively different from that in standard diffusive metals. This distinction shows up most dramatically in the FCS, which we have studied, but also has implications for the nature of nonlinear response and nonequilibrium charge noise~\cite{natelson}, which we will address in future work.

{\it Acknowledgements.} We thank Luca Delacr\'etaz,  Nathalie de Leon, Jacopo De Nardis, Matthew Foster, Paolo Glorioso, David Huse, Vedika Khemani, Andrew Lucas, Alan Morningstar, and Frank Zhang for helpful discussions, collaboration on previous works  and/or comments on this manuscript.
We acknowledge support from NSF Grants No. DMR-2103938 (S.G., E.M.) and DMR-2104141 (E.M., R.V.).

\bibliography{Refs}
\bibliographystyle{apsrev4-1}

\end{document}